\documentclass[10pt,letterpaper,twocolumn,showpacs,prl,aps]{revtex4-1} 

\usepackage{amsmath}  
\usepackage{amsfonts} 
\usepackage{graphicx} 
\usepackage{amssymb}
\usepackage{url}
\usepackage{hyperref}

\begin{document}

\title{Nonlinear Meissner States, Vortex Sheets, and Laminar Structures in Extreme Type-II Superconductors}

\author{Eugene B. Kolomeisky}

\affiliation
{Department of Physics, University of Virginia, P. O. Box 400714, Charlottesville, Virginia 22904-4714, USA}

\date{\today}

\begin{abstract}
A recently derived nonlinear velocity theory of extreme type-II superconductors is shown to possess an exactly solvable one-dimensional sector. Its solutions include nonlinear Meissner states exhibiting universal tails, vortex sheets, and periodic laminar structures. The thermodynamic critical field emerges from a marginal Meissner profile equivalent to the normal-superconducting boundary and may be viewed as a half-soliton. The vortex sheet is a soliton characterized by a discontinuity of the velocity field, a continuous and localized magnetic field and vanishing superconducting density at its center, and may be interpreted as the coarse-grained limit of a dense row of Abrikosov vortices. Periodic solutions describe laminar states that may be interpreted as coarse-grained rectangular vortex lattices. 
\end{abstract}

\maketitle

The Ginzburg-Landau (GL) theory \cite{GL} provides a remarkably successful description of superconductivity over a broad range of length scales and magnetic fields. In the limit of extreme type-II superconductors, characterized by a large GL parameter $\kappa=\lambda/\xi$, the ratio of the London penetration depth to the coherence length, the theory is commonly viewed as reducing to the London description outside narrow vortex cores \cite{Abrikosov1,Abrikosov2,de Gennes,LL9,Fetter,Sarma}. This expectation has recently been challenged by the derivation of a nonlinear velocity theory obtained directly from the GL theory in the limit $\kappa\rightarrow\infty$ \cite{EBK}. Despite involving a single length scale, the resulting theory remains intrinsically nonlinear and admits nontrivial exact solutions.

A particularly striking consequence of the nonlinear velocity theory is the existence of an exactly solvable Abrikosov vortex whose outer structure differs qualitatively from the conventional London picture. This observation suggests that the extreme type-II limit does not merely simplify the GL theory but reveals a previously hidden nonlinear sector. An important question is whether similarly nontrivial structures exist beyond the isolated vortex and, in particular, whether the reduced theory admits exact solutions relevant to screening and mixed-state phenomena.

In this Letter we investigate the one-dimensional transverse sector of the nonlinear velocity theory. We show that it contains exact nonlinear Meissner states exhibing universal tails, normal-superconducting (NS) interfaces, solitons, and periodic structures. The thermodynamic critical field emerges from a marginal Meissner state.  The NS boundary is the marginal Meissner profile representing a half-soliton. The full soliton is a vortex sheet characterized by a discontinuity of the velocity field, vanishing superconducting density, and a continuous and localized magnetic field. We argue that this object may be interpreted as the coarse-grained limit of a dense row of Abrikosov vortices, thus providing a connection between the reduced theory and finite values of $\kappa$.

The theory also admits periodic solutions describing laminar structures that may be viewed as coarse-grained rectangular vortex lattices, establishing a unified description of nonlinear screening, vortex sheets, and mixed-state profiles within a simple exactly solvable framework. The results further demonstrate that the limit $\kappa\rightarrow\infty$ retains a rich nonlinear content and provides a novel perspective on extreme type-II superconductivity.

Adopting the GL units \cite{GL,Abrikosov1,Abrikosov2,LL9,Fetter}, in the $\kappa\to\infty$ limit, the GL free energy relative to the reference superconducting state reduces to the functional of the superfluid velocity $\mathbf{v}$ only \cite{EBK}
\begin{equation}
\label{AL_free_energy}
F[\mathbf{v}]=\int \left [(\nabla\times \mathbf{v})^{2}+\mathbf{v}^{2}-\frac{1}{2}\mathbf{v}^{4}\right ]dV,~~|\mathbf{v}|\le 1
\end{equation}
where the constraint $|\mathbf{v}|\le 1$ follows from the condition
\begin{equation}
\label{slaving_condition}
\rho=1-\mathbf{v}^{2}
\end{equation}
relating the density of superconducting electrons $\rho$ to the velocity $\mathbf{v}$.  The maximal velocity $v_{L}=1$ is the Landau critical velocity \cite{LL9} of the $\kappa\to\infty$ theory.  Within the reduced velocity theory, $\mathbf{v}^{2}=1$ represents the normal, $\rho=0$, state from the superconducting side of the theory.  

Since the free energy density in (\ref{AL_free_energy}) contains the magnetic induction $\mathbf{b}=-\nabla\times\mathbf{v}$, nonuniform solutions minimizing the functional (\ref{AL_free_energy}) are transverse, $\nabla\mathbf{v}=0$.  One such solution is the Abrikosov vortex
\begin{equation}
\label{Abrikosov_vortex}
v_{\varphi}=\frac{1}{\kappa}K_{1}(r),~~~~ b=\frac{1}{\kappa}K_{0}(r),~~~~\rho=1-\frac{K_{1}^{2}(r)}{\kappa^{2}}
\end{equation} 
whose axis is the $z$-axis of the cylindrical coordinate system, $r$ is the distance from the axis, $v_{\varphi}$ is the velocity pointing in the azimuthal direction, $b$ is the the magnetic induction pointing in the $z$-direction, and $K_{\nu}(r)$ is the modified Bessel function of the second kind of order $\nu$ \cite{NIST}.  Even though Eq.(\ref{AL_free_energy}) is parameter-free, the GL parameter $\kappa$ reappears in the solution (\ref{Abrikosov_vortex}) due to the singular boundary condition $v_{\varphi}(r\to 0)=1/\kappa r$ ensuring quantization of the magnetic flux \cite{Abrikosov1,Abrikosov2}.  Eq.(\ref{Abrikosov_vortex}) is asymptotically exact, $\kappa\to\infty$, solution of the Abrikosov vortex outside the vanishing normal vortex core of size $\xi=1/\kappa$, the coherence length \cite{EBK}.           

The one-dimensional transverse sector of the velocity theory corresponds to the choice $\mathbf{v}=\left(0, v(x), 0\right)$ and $\mathbf{b}=(0,0,b(x))$ where $b=-dv/dx\equiv -v'$.  Accounting for an external magnetic field $h$ in the positive $z$ direction, the Gibbs free energy \cite{GL,de Gennes,Abrikosov1,Abrikosov2,LL9,Fetter,Sarma} per unit area is given by
\begin{eqnarray}
\label{free_energy_per_unit area}
g[v]&=&2\int\left (\frac{v'^{2}}{2}+ U(v)+hv'\right )dx,\nonumber\\
U(v)&=&\frac{1}{2}v^{2}-\frac{1}{4}v^{4},~~~~|v|\le 1.
\end{eqnarray}
Minimizing the functional $g[v]$ with respect to $v$ yields
\begin{equation}
\label{Duffing}
v''=\frac{dU}{dv}=j=\rho v=(1-v^{2})v
\end{equation} 
which is Amp\`ere's law with the current density $j$.  Equation (\ref{Duffing}) serves as the universal one-dimensional field equation of the nonlinear velocity theory.

The character of solutions to Eq.(\ref{Duffing}) can be understood based on a mechanical analogy:  it can be viewed as Newton's second law of motion for a fictitious particle of unit mass and position $v$ moving in the field of the potential energy $-U(v)$; the role of time is played by $x$.  The field equation (\ref{Duffing}) has the form of the Duffing equation \cite{Duffing,Nayfeh}, although the physical solutions are restricted by the superconducting constraint $|v|\le 1$ and thermodynamics.  The separatrix solution of the mechanical problem (\ref{Duffing}) corresponding to its first integral
\begin{equation}
\label{energy_conservation}
v'^{2}\equiv b^{2}=2U(v),
\end{equation}
is spatially localized and connects the two maxima of $U(v)$. In the language of nonlinear field theory, it represents a static soliton. In the present context, it interpolates between the superconducting state, $v=0$, and the normal state, $v^{2}=1$.

The nonlinear Meissner effect provides the first physical application of the soliton solution.  Near the superconducting state ($v=0$), Eq.(\ref{Duffing}) linearizes, giving $v\simeq \exp(\pm x)$, so that the magnetic induction decays exponentially into the bulk.  At the vacuum-superconductor boundary $x=x_{s}$, the condition $b(x_{s})=-v'=h$ yields
\begin{equation}
\label{critical_field}
h^{2}=2U[v(x_{s})]\le 2U(v_{L})=\frac{1}{2}.
\end{equation}    
Thus, the Meissner state exists only for $h\le h_{c}=1/\sqrt{2}$ where $h_{c}^{2}$ equals the free-energy density difference between the normal ($v^{2}=1$) and superconducting ($v=0$) states, as required thermodynamically \cite{GL,Abrikosov1,Abrikosov2,Fetter,Sarma,LL9}. In the velocity theory this finite critical field follows directly from the bounded potential $U(v)$, whose maximum occurs at the Landau critical velocity $v=\pm v_{L}=\pm 1$.  Thus, the thermodynamic critical field $h_{c}$ emerges as the field counterpart of the Landau critical velocity. By contrast, the London theory corresponds to $U(v)=v^{2}/2$, for which no finite critical field exists.  

\begin{figure}
\begin{center}
\includegraphics[width=1\columnwidth]{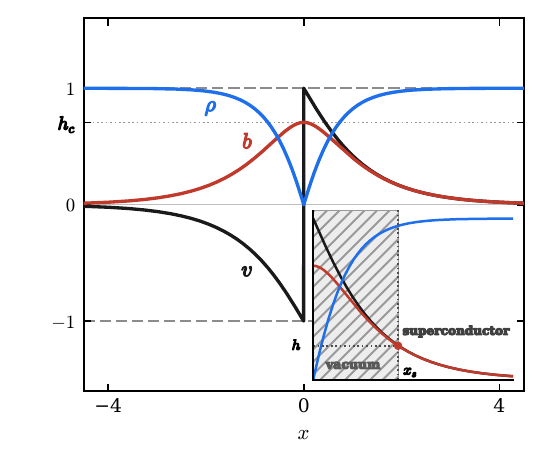}
\caption{The velocity $v$, magnetic induction $b$ and superconducting density $\rho$ in the soliton (\ref{velocity}).  Inset: Construction of Meissner states by truncating the universal soliton at the vacuum-superconductor boundary $x=x_{s}$.}
\label{soliton}
\end{center}
\end{figure}

Equation (\ref{energy_conservation}) can be solved yielding
\begin{eqnarray}
\label{velocity}
v&=&\pm\frac{\sqrt{2}}{\cosh(x+x_{0})},~~~~~~x_{0}=\ln(\sqrt{2}\pm1),\nonumber\\
b&=&\pm\frac{\sqrt{2}\sinh(x+x_{0})}{\cosh^{2}(x+x_{0})},~~\rho=1-\frac{2}{\cosh^{2}(x+x_{0})}
\end{eqnarray}
where the upper (lower) sign applies for positive (negative) $x$, the origin has been chosen so that $|v(0)|=1$ and the expressions for the magnetic induction $b=-v'$ and superconducting density (\ref{slaving_condition}) are also given.  These three quantities constitute the universal soliton profile shown in Figure \ref{soliton}

The physical significance of the soliton is that it encompasses the entire family of nonlinear Meissner states. Consider a superconducting half-space bounded by a vacuum at $x=x_{s}\ge 0$ as shown in the inset in Figure \ref{soliton}.  The boundary condition $b(x_{s})=h$ 
uniquely determines the boundary position $x_{s}(h)$, so that the Meissner state is simply the part of the soliton with $x\ge x_{s}(h)$. Thus every nonlinear Meissner profile is a truncated half-soliton.  An immediate consequence is the universality of the Meissner tails.   Different external fields merely select different truncation points on the same underlying profile, while the asymptotic region remains unchanged. In particular,
\begin{equation}
\label{tail}
b[x\gg x_{s}(h)]=2(2-\sqrt{2})\exp(-x)\approx 1.17 \exp(-x)
\end{equation}
independently of the applied field.  Although the exponential decay agrees with the London theory \cite{de Gennes,LL9}, the amplitude does not scale with the external field. Thus the nonlinear velocity theory predicts universal magnetic tails even in the low-field regime usually regarded as the domain of applicability of the London theory \cite{de Gennes}.  As the applied field increases, the vacuum-superconductor boundary $x_{s}(h)$, moves monotonically toward the soliton center, reaching $x_{s}=0$ at the thermodynamic critical field $h=h_{c}=1/\sqrt{2}$.  The corresponding marginal Meissner state is therefore precisely the half-soliton. No Meissner solution exists for larger fields.

For the marginal Meissner profile, the velocity, magnetic induction, and superconducting density connect the superconducting state ($v=0$, $b=0$, $\rho=1$) to the normal state ($v=1$, $b=h_{c}=1/\sqrt{2}$, $\rho=0$). These values correspond to the endpoint of the superconducting branch of the NS boundary and, at $h=h_{c}$ coincide with the thermodynamically degenerate normal phase.  At $h=h_{c}$, the Gibbs free-energy densities of the normal and superconducting phases are equal, so the NS surface tension arises solely from the interfacial (superconducting) region.

The physical interpretation of the soliton, Figure \ref{soliton}, follows from a simple limiting construction. Consider two NS boundaries at the thermodynamic critical field, enclosing a normal slab of finite width. Since the magnetic induction is everywhere equal to the thermodynamic critical field within the normal region, the magnetic fields associated with the two boundaries do not overlap, and the excess free energy is therefore exactly twice the NS surface tension. As the width of the normal slab is continuously reduced to zero, the two boundaries merge into a single object, exhibiting a tangential discontinuity of the velocity.  This is the soliton or vortex sheet.  No measurable singularity develops in this limit: the superconducting density already vanishes where the boundaries meet, so the current density $j=\rho v=(1-v^{2})v$ is identically zero at the sheet. Consequently, the magnetic induction remains continuous and finite, and the energy of the limiting configuration is twice the NS surface energy. The vortex sheet may thus be viewed as two NS boundaries placed back-to-back, while the NS boundary itself is naturally interpreted as a half-soliton.   

Curiously, in the Onsager-Landau-Lifshitz picture of rotating superfluids \cite{Onsager,LL55,Volovik}, vortex sheets were proposed as a mechanism to locally destroy superfluidity and accommodate rotation. In the present theory, the soliton is interpreted as the superconducting counterpart of a vortex sheet, locally destroying the superconducting state.

Employing Eqs.(\ref{free_energy_per_unit area}) and (\ref{energy_conservation}), the surface energy of the soliton in an external field $h$ is
\begin{eqnarray}
\label{soliton_surface_energy}
\sigma(h)&=&4\left (\int_{0}^{\infty}b^{2}dx-h\right )=4\left (\int_{0}^{1}\sqrt{2U(v)}dv-h\right )\nonumber\\
&\equiv&4(h_{c_{1}}-h),~~~h_{c_{1}}=\frac{2}{3}\left (1-\frac{1}{2\sqrt{2}}\right )\approx 0.43.
\end{eqnarray}
The second representation expresses the result entirely in terms of the potential $U(v)$, without requiring the explicit soliton profile. At the thermodynamic critical field, the soliton may be viewed as two back-to-back NS boundaries, yielding the surface tension $\sigma_{NS}=\sigma(h_c)/2=-4(\sqrt{2}-1)/3<0$.  An equivalent integral representation was employed by Saint-James, Sarma, and Thomas \cite{Sarma} to calculate the surface tension of the NS boundary within the GL theory. In the present work, it emerges naturally from the nonlinear velocity theory and is extended to the field-dependent surface energy of the soliton.

Vortex sheets become thermodynamically favorable when the soliton surface energy (\ref{soliton_surface_energy}), positive at $h=0$, changes sign which occurs at $h=h_{c_{1}}$.
Above $h_{c_{1}}<h_{c}=1/\sqrt{2}\approx0.71$, the equilibrium density of vortex sheets increases with the applied field, resulting in progressively greater magnetic field penetration and a laminar mixed state. In this respect, the vanishing of the soliton energy at $h_{c_{1}}$ is directly analogous to the onset of proliferation of Abrikosov vortices at their lower critical field. Since the latter is $O(1/\kappa)$ \cite{Abrikosov1,Abrikosov2}, vortex lattices remain thermodynamically preferred in sufficiently weak fields. Nevertheless, vortex sheets may become competitive at higher fields or in geometries possessing planar symmetry, where they are naturally accommodated.

A laminar mixed state in type-II superconductors was proposed by Goodman \cite{Goodman,de Gennes,Sarma}, motivated by the layered structure of the intermediate state in type-I superconductors and analyzed within the London theory. The nonlinear velocity theory provides an exact description of the elementary lamina as a vortex-sheet solution (\ref{velocity}) from which the properties of the laminar state discussed below follow directly.

There is a close correspondence between the vortex sheet (\ref{velocity}) and the Abrikosov vortex (\ref{Abrikosov_vortex}). The latter is characterized by quantized velocity circulation responsible for flux quantization. In one dimension, the corresponding quantity is the velocity discontinuity (circulation per unit length), which determines the magnetic flux per unit length carried by the vortex sheet:  $\int_{-\infty}^{\infty}bdx=-\int v'dx=v(0+)-v(0-)=2$.  Moreover, in the limit $\kappa\to\infty$, the tangential velocity of the Abrikosov vortex (\ref{Abrikosov_vortex}) undergoes the same jump of magnitude $2$ across the vortex axis, establishing a direct correspondence with the vortex sheet.

The nonlinear velocity theory is macroscopic with respect to the GL theory, in the same sense that hydrodynamics is macroscopic with respect to molecular dynamics.  The nonlinear velocity theory is obtained by coarse-graining the GL theory over distances of order $1/\kappa$. Consequently, structures varying on this scale are not resolved and may appear macroscopically as discontinuities.  This is closely analogous to hydrodynamics, where a vortex sheet is the continuum limit of a dense row of Euler vortices \cite{Lamb}.  We therefore interpret the superconducting vortex sheet as the coarse-grained description of a dense row of Abrikosov vortices with spacing $O(1/\kappa)$.

This interpretation is supported by the scaling of the magnetic induction.  The field at the center of an isolated Abrikosov vortex is $O(\ln\kappa/\kappa)$ (\ref{Abrikosov_vortex}), whereas the vortex sheet has a field of order unity (\ref{velocity}). A row with spacing $O(1/\kappa)$ contains $O(\kappa)$ vortices within one London penetration depth. Their collective contribution is therefore $O(\ln\kappa)$, which, up to the logarithmic factor neglected in this scaling argument, is consistent with the thermodynamic critical field. 

Having exhausted the separatrix solution of Eq. (\ref{Duffing}), we now turn to its periodic solutions described by the first integral
\begin{equation}
\label{energy_conservation_laminar}
v'^{2}\equiv b^{2}=b_{0}^{2}+2U(v)
\end{equation}
which corresponds to laminar mixed states.  The integration constant $b_{0}$ is the minimum magnetic induction attained where $v=0$ and $\rho=1$.  The maximum induction $b_{max}=\sqrt{b_{0}^{2}+1/2}$ occurs where  $v^{2}=1$ and $\rho=0$.  As $b_{0}$ increases, the modulation of the magnetic induction decreases, so that $b_{0}(h)$ increases monotonically from $b_{0}(h_{c_{1}})=0$ to $b_{0}(h)\to h$ as $h\to\infty$.  

Eq.(\ref{energy_conservation_laminar}) can be integrated yielding 
\begin{equation}
\label{quadrature}
x=-\int\frac{dv}{\sqrt{b_{0}^{2}+2U(v)}}+const.
\end{equation}
Within each period of length
\begin{equation}
\label{period}
\Lambda(b_{0}^{2})=2\int_{0}^{1}\frac{dv}{\sqrt{b_{0}^{2}+2U(v)}},
\end{equation}
$v(x)$ varies between $-1$ and $1$, and periodic continuation produces a laminar array of vortex sheets separated by superconducting regions. Each sheet exhibits the same velocity jump of magnitude $2$ as the isolated soliton. In the limit $b_{0}\to 0$, the period diverges and the solution reduces to a train of well-separated solitons (\ref{velocity}).

The period (\ref{period}) also determines the average magnetic induction
\begin{equation}
\label{average_induction}
\overline{b}=\frac{1}{\Lambda}\int_{\Lambda}b(x)dx=-\frac{1}{\Lambda}\int_{\Lambda}v'(x)dx=\frac{2}{\Lambda}
\end{equation}
where the integration is performed over any interval of length $\Lambda$.  Thus each period contributes the same flux per unit length, equal to the velocity jump of the vortex sheet.  Proceeding as for the isolated soliton (see Eq.(\ref{soliton_surface_energy})), the Gibbs free energy density  of a laminar state becomes
\begin{eqnarray}
\label{Gibbs_free_energy_density}
\frac{g}{\Lambda}&=&\frac{4}{\Lambda}\left (\int_{0}^{1}\sqrt{b_{0}^{2}+2U(v)}dv-h\right )-b_{0}^{2}\nonumber\\
&=&\frac{4}{\Lambda}(h_{c_{1}}-h)+\frac{1}{\Lambda}\int_{0}^{b_{0}^{2}}\Lambda(y)dy-b_{0}^{2}.
\end{eqnarray}
Minimization of Eq. (\ref{Gibbs_free_energy_density}) with respect to $b_{0}$ or equivalently $\Lambda$ or the average induction $\overline{b}$, determines the equilibrium laminar state as a function of the applied field.  

Equation (\ref{quadrature}) is expressible in terms of elliptic integrals, and its inverse in terms of Jacobi elliptic functions, as is characteristic of equations of Duffing type (\ref{Duffing}). Since these expressions merely interpolate smoothly between the asymptotic regimes discussed below, we do not display them explicitly.

In the dilute limit $b_{0}\ll1$, the quartic term in the potential $U(v)$ (\ref{Duffing}) may be neglected, giving for the period (\ref{period}) $\Lambda(y)=\ln(4/y)$.  The resulting Gibbs free energy density (\ref{Gibbs_free_energy_density})
\begin{equation}
\label{CIC_Gibbs_free_energy}
\frac{g}{\Lambda}=\frac{4}{\Lambda}(h_{c_{1}}-h)+\frac{4}{\Lambda}e^{-\Lambda}
\end{equation}
admits a simple interpretation in terms of interacting solitons.  The first term is simply the soliton energy density;  its surface energy counterpart was already found earlier (\ref{soliton_surface_energy}).  The second term decays exponentially with the soliton separation $\Lambda$ and therefore represents an exponentially weak repulsive interaction between neighboring solitons.  Minimization with respect to the soliton density (or equivalently the average induction) yields
\begin{equation}
\label{h_vs_b}
h-h_{c_{1}}=\Lambda e^{-\Lambda}=\frac{2}{\overline{b}}e^{-2/\overline{b}}
\end{equation}  
which determines the $\overline{b}(h)$ dependence.   Specifically, neglecting higher order terms we find 
\begin{equation}
\label{susceptibility}
\overline{b}\simeq 2\ln^{-1}\frac{1}{h-h_{c_{1}}},~~\Lambda=\frac{2}{\overline{b}}=2\ln\frac{2}{b_{0}}\simeq \ln\frac{1}{h-h_{c_{1}}}.
\end{equation}
Thus the average induction increases only logarithmically above the lower critical field.  It is worth noting that as $h\to h_{c_{1}}$ the derivative $d\overline{b}/dh$ diverges according to $d\overline{b}/dh\propto (h-h_{c_{1}})^{-1}\ln^{-2}[1/(h-h_{c_{1}})]$.  The logarithmic dependences (\ref{susceptibility}) are characteristic of zero-temperature commensurate-incommensurate transitions \cite{Bak}.

Increasing the external field reduces the separation between neighboring solitons. The dilute-soliton description ceases to be applicable once their spacing becomes comparable to the London penetration depth, $\Lambda=O(1)$. This occurs for $h-h_{c_{1}}=O(1)$, the thermodynamic critical field.   At this crossover, $\overline{b}\simeq b_{0}=O(1)$, and the soliton train evolves smoothly into a denser laminar array of vortex sheets.

Since each vortex sheet is interpreted as the coarse-grained limit of a dense row of Abrikosov vortices, the laminar state may be viewed as the macroscopic description of a rectangular Abrikosov vortex lattice.

In the dense limit $b_{0}\gg 1$, the potential $U(v)$ (\ref{Duffing}) may be neglected, giving for the period (\ref{period}) $\Lambda(y)=2/\sqrt{y}$.  The Gibbs free energy density (\ref{Gibbs_free_energy_density}) assumes the particularly simple form          
\begin{equation}
\label{large_field_Gibbs_free_energy}
\frac{g}{\Lambda}=\frac{4}{\Lambda^{2}}-\frac{4h}{\Lambda}=\overline{b}^{2}-2\overline{b}h
\end{equation}
whose minimization yields
\begin{equation}
\label{large_field_equilibrium}
\overline{b}=b_{0}=h,~v=-hx,~ \rho=1-h^{2}x^{2},~ |x|\le \frac{\Lambda}{2}=\frac{1}{h}.
\end{equation}
Despite full penetration of the magnetic field into the superconductor, there remain substantial variations of the velocity and superconducting density that we also displayed.  The period of the structure continues to decrease with an increase of the external field until the field reaches the value $O(\kappa)$.   At  that point the period $\Lambda=O(1/\kappa)$ attains the resolution limit of the velocity theory.

Since laminar structures may be viewed as coarse-grained rectangular vortex lattices, at the field $O(\kappa)$ the distance between neighboring vortices becomes $O(1/\kappa)$, and a transition to the normal state is expected to occur.  This is consistent with the upper critical field of transition of the Abrikosov lattice into the normal state to be precisely $h_{c_{2}}=\kappa$ \cite{Abrikosov1,Abrikosov2,de Gennes,LL9}.

Interestingly, the average values $\overline{v^{2}}=1/3$ and $\overline{\rho}=2/3$, inferred from Eqs.(\ref{large_field_equilibrium}), coincide with those characterizing the depairing current of a thin wire \cite{Abrikosov2} which is the maximum of the current-velocity relation $j=\rho v=(1-v^{2})v$ in the field equation (\ref{Duffing}). Since both results originate from the same cubic in $v$ nonlinearity, this correspondence is not accidental.  For the wire the density of superconducting electrons drops abruptly from $\rho=2/3$ to $\rho=0$ at the depairing current.  These observations suggests that the large-field laminar state may lie on the verge of a first-order transition, a possibility that deserves further investigation.

Rather than introducing new physics, the nonlinear velocity theory reveals physical structures already contained within the underlying Ginzburg-Landau theory. It demonstrates that the extreme type-II limit is not a trivial asymptotic reduction but a nontrivial field theory in its own right. Hidden by the two-scale character of the original formulation, nonlinear Meissner states, vortex sheets, and laminar mixed states emerge naturally while remaining analytically tractable. These results suggest that the $\kappa\to\infty$ limit provides a natural starting point for understanding superconductivity in extreme type-II materials, with finite-$\kappa$ effects entering as systematic corrections.

\end{document}